\documentclass[reprint,aps,pre,amsmath,amssymb,showpacs,nofootinbib,onecolumn,notitlepage]{revtex4-1}
\usepackage[pdftex]{graphicx}	
\usepackage{xcolor}
\graphicspath{ {figs/} }

\usepackage{hyperref}
\hypersetup{
	breaklinks=true,
	colorlinks=true,
	linkcolor=blue,
	citecolor=purple,
	pdfstartview={FitH},
	pdfview={FitH 0},
	pdfauthor={Jeffrey B. Parker and Navid C. Constantinou},
	pdftitle={Magnetic eddy viscosity of mean shear flows in two-dimensional magnetohydrodynamics},
 }

\usepackage{bm}
\usepackage{eucal}		
\usepackage{mycommands}   

\newcommand{\Alfven}{Alfv\'en}

\newcommand{\tc}{\t_{\rm corr}}  
\newcommand{\Lt}{L_{\rm turb}}
\newcommand{\vt}{v_{\rm turb}}
\renewcommand{\tt}{\t_{\rm turb}}

\newcommand{\Rm}{\mathrm{Rm}}

\newcommand{\mathtensor}[1]{\bm{\mathsf{#1}}}

\providecommand{\eqnref}{}
\renewcommand{\eqnref}[1]{Eq.~\eqref{#1}}

\newcommand{\drag}{\kappa}
\newcommand{\Btilde}{\tilde{B}}    

\begin{document}

\title{Magnetic eddy viscosity of mean shear flows in two-dimensional magnetohydrodynamics}

\author{Jeffrey B.~Parker}
\email{parker68@llnl.gov}
\affiliation{Lawrence Livermore National Laboratory, Livermore, CA 94550}

\author{Navid C. Constantinou}
\affiliation{Research School of Earth Sciences and ARC Centre of Excellence for Climate Extremes, Australian National University, Canberra, ACT, 2601, Australia}

\begin{abstract}
Magnetic induction in magnetohydrodynamic fluids at magnetic Reynolds number (Rm) less than~1 has long been known to cause magnetic drag.  Here, we show that when $\Rm \gg 1$ and the fluid is in a hydrodynamic-dominated regime in which the magnetic energy is much smaller than the kinetic energy, induction due to a mean shear flow leads to a magnetic eddy viscosity.  The magnetic viscosity is derived from simple physical arguments, where a coherent response due to shear flow builds up in the magnetic field until decorrelated by turbulent motion.  The dynamic viscosity coefficient is approximately $(B_p^2/2\m_0) \tc$, the poloidal magnetic energy density multiplied by the correlation time.  We confirm the magnetic eddy viscosity through numerical simulations of two-dimensional incompressible magnetohydrodynamics.  We also consider the three-dimensional case, and in cylindrical or spherical geometry, theoretical considerations similarly point to a nonzero viscosity whenever there is differential rotation.  Hence, these results serve as a dynamical generalization of Ferraro's law of isorotation.  The magnetic eddy viscosity leads to transport of angular momentum and may be of importance to zonal flows in astrophysical domains such as the interior of some gas giants.
\end{abstract}

\maketitle

\section{Introduction}
\label{sec:intro}
The combination of rotation and magnetic fields is common in astrophysical fluid dynamics, found in stellar interiors, gas giant atmospheres, and astrophysical disks.  In the outer atmosphere of planets, magnetic fields do not play a significant role; the flow is dominated by rotation, which leads to anisotropic flow in which turbulence, waves, and coherent structures interact.  A common feature is the presence of long-lived mean shear flows, called zonal flow, which alternate in latitude.  In Jupiter's atmosphere, for instance, zonal flows are remarkably persistent \cite{vasavada:2005}.  However, deeper into gas giant interiors, or in exoplanets, magnetic fields may play an important role and influence the flow.

In resistive magnetohydrodynamics (MHD), it is well known that when the magnetic Reynolds number is less than one, fluid flow across a magnetic field results in magnetic drag \cite{davidson:1995}.  The magnetic Reynolds number $\Rm = LV / \eta$ measures the relative strength of induction and magnetic diffusion, where $L$ is a characteristic length scale of the system, $V$ is a characteristic fluid velocity, and $\eta$ is the magnetic diffusivity.

The regime $\Rm \gg 1$ is more dynamical and complex.  Within this regime we examine a subset of parameter space: we assume large Reynolds number and consequently turbulent flow.  We also assume a hydrodynamic-dominated regime where the influence of the magnetic field is small.  We characterize this latter restriction with the dimensionless parameter $\mathcal{A} \ll 1$, where $\mathcal{A} = |\v{J} \times \v{B}| / |\r \v{v} \cdot \v{\nabla} \v{v}|$ is the ratio of the Lorentz force to the inertial force.  If the energy-containing scale lengths of the magnetic field and velocity field are comparable, then equivalently $\mathcal{A} = (B^2 / \m_0) / (\r v^2)$, i.e., the ratio of magnetic energy density to kinetic energy density.  Here, $\v{J}$ is the current density, $\v{B}$ the magnetic field, $\v{v}$ the fluid velocity, $\r$ the fluid mass density, and $\m_0$ the permeability of free space.

\begin{figure}
		\includegraphics{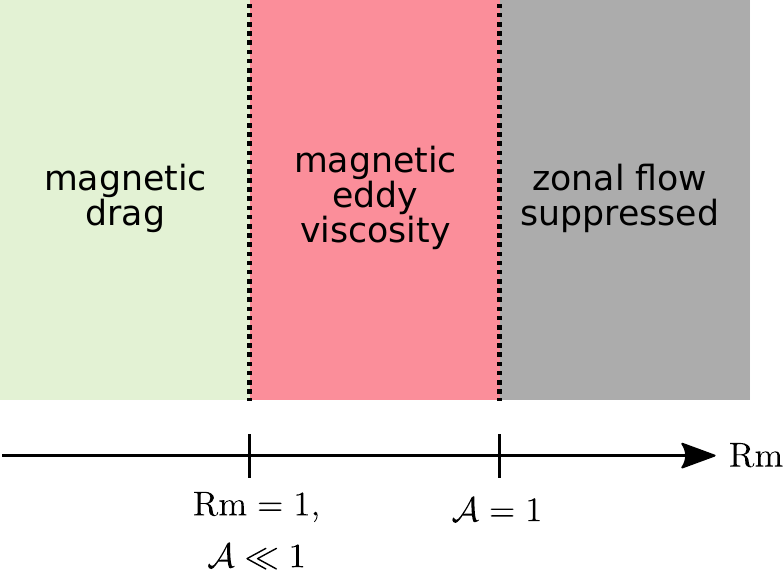}
		\caption{Schematic showing the magnetic response to a mean shear flow expected in a turbulent regime.  As $\Rm$ increases, the magnetic response transitions from a magnetic drag (for flow perpendicular to the magnetic field) for $\Rm < 1$ to a magnetic eddy viscosity for $\Rm > 1$, assuming also $\mathcal{A} \ll 1$.  As $\Rm$ increases further, the magnetic energy rises to a value comparable to the kinetic energy, $\mathcal{A}=1$.  If the mean shear flow is turbulence-driven, such as zonal flow, then the mean flow is suppressed.}
		\label{fig:parameterspace}
	\end{figure}

In this paper, we suggest that within the above regime, the Lorentz force on mean shear flow acts not as a magnetic drag but rather as an effective magnetic eddy viscosity.  We demonstrate this result in two-dimensional (2D) incompressible MHD simulations.  Under conditions where $\mathcal{A}$ approaches unity, the Lorentz force becomes appreciable.  If the mean shear flow is a turbulence-driven zonal flow, the magnetic eddy viscosity can be strong enough to suppress it.  Figure~\ref{fig:parameterspace} offers a schematic diagram showing the parameter space within which the magnetic eddy viscosity is expected in a turbulent flow.

This paper is primarily a study of MHD physics in an idealized setup; however, a potential example where the regime $\Rm \gg 1$ and $\mathcal{A} \ll 1$ could be realized is in the interiors of gas giants.  In the outer atmospheres of Jupiter and Saturn, the bulk fluid is electrically insulating, but the conductivity increases rapidly with depth, giving way to a conducting fluid in the interior~\cite{french:2012,liu:2008}.  Recent spacecraft measurements have shed light on long-standing questions regarding the depth of zonal flow in planetary interiors.  Gravitometric observations indicate that the zonal flows in Jupiter and Saturn terminate at a depth of approximately 3,000 and 8,500 km, respectively \cite{kaspi:2018,guillot:2018,iess:2019,galanti:2019}.  These depths are near the transition zone where the electrical conductivity increases rapidly.  The greatly enhanced conductivity in the interior leads to a stronger influence of the magnetic field on bulk fluid motion.  The coincident depths of the suppression of zonal flow and the rise in conductivity in Jupiter and Saturn have led to the hypothesis that the magnetic field is responsible for terminating differential rotation.  Moreover, simulations of planetary dynamo models provide some support for this possibility \cite{heimpel:2011,duarte:2013,jones:2014,dietrich:2018}.  While magnetic drag at $\Rm < 1$ has been studied for gas giants and hot Jupiters as the mechanism to suppress differential rotation \cite{cao:2017, perna:2010}, there remains a great deal of uncertainty about parameter values.  Intermediate regimes may exist in which $\Rm \gg 1$ and  $\mathcal{A} \ll 1$. 

Idealized settings to study individual effects prove fruitful.  For example, certain aspects of fluid physics under the influence of rotation and magnetization can be studied in the 2D $\b$ plane or 2D spherical surface \cite{tobias:2007,tobias:2011,constantinou:2018,durston:2016,zinyakov:2018}.  A $\b$ plane is a Cartesian geometrical simplification of a rotating sphere that retains the physics associated with rotation and the latitudinal variation of the Coriolis effect \citep{pedlosky:book}.  The unmagnetized $\b$ plane is one of the simplest settings in which coherent zonal flows emerge spontaneously from turbulence and is often used for idealized studies \cite{rhines:1975, vallis:1993}.  Such idealizations eliminate many sources of complexity, including thermal convection, variation in the density and electrical conductivity, and true 3D dynamics.  Two-dimensional MHD studies have sufficed, however, to demonstrate magnetic suppression of zonal flow \cite{tobias:2007,tobias:2011,constantinou:2018}.  Here, our numerical simulations continue with the 2D magnetized $\b$ plane, although our physical considerations are quite general.  The magnetic eddy viscosity also sheds light on earlier results on the magnetized $\b$ plane and magnetized 2D spherical surface.

An outline for the rest of this paper is as follows.  In Sec.~\ref{sec:magneticviscosity}, we derive the magnetic eddy viscosity of mean shear flow.  In Sec.~\ref{sec:negative_eddy_viscosity}, we recall the negative eddy viscosity for comparison.  In Sec.~\ref{sec:simulation_setup}, we describe the setup for our 2D simulations.  The simulation results are shown in Sect.~\ref{sec:simulation_results}.  Section~\ref{sec:3d_generalization} provides the 3D generalization of the magnetic eddy viscosity.  Section~\ref{sec:planet} applies these results to consider zonal flows deep in planetary interiors.  We summarize our results in Sec.~\ref{sec:conclusion}.



\section{Magnetic eddy viscosity}
\label{sec:magneticviscosity}
In this section we consider a mean shear flow $\v{v} = U(y) \unit{x}$ and investigate the zonally averaged magnetic force acting back on the flow due to the correlations induced by the flow.  The zonal average of a quantity $f$ is defined as
	\begin{equation}
		\ol{f} \defineas \frac{1}{L_x} \int_0^{L_x} dx\, f,
	\end{equation}
where $L_x$ is the length of the domain in the $x$ direction, assumed to be periodic.  The deviation from the zonal mean is denoted by a prime, i.e.,
	\begin{equation}
		f' \defineas f - \ol{f}.
	\end{equation}

The Lorentz force $\v{F} = \v{J} \times \v{B}$ acting on a fluid can be expressed as $\v{\nabla} \cdot \mathtensor{T}$, where the Maxwell stress tensor~$\mathtensor{T}$ is
	\begin{equation}
		\mathtensor{T} = \frac{\v{B}\v{B}}{\m_0} - \frac{B^2}{2\m_0} \mathtensor{I},
	\end{equation}
and the stress terms associated with the electric field are neglected within the MHD approximation.  This formulation of the Lorentz force will be most convenient for our purposes, though we return to consideration of the current in Appendix~\ref{appendix:force_from_current}.  The force per unit volume in the $x$ direction is
	\begin{equation}
		F_x = \pd{T_{xx}}{x} + \pd{T_{yx}}{y} + \pd{T_{zx}}{z}.
	\end{equation}
The first term on the right-hand side (RHS) vanishes after a zonal average.  In the remainder of this section, we assume a quasi-2D system and set $B_z=0$ and $\partial_z=0$, so
	\begin{equation}
		\ol{F}_x = \pd{\ol{T}_{yx}}{y} = \pd{}{y} \frac{\ol{B_x B_y}}{\m_0}.
	\end{equation}
	
We evaluate the coherent effect that a shear flow has on inducing correlations in the magnetic field and in the corresponding backreaction on the shear flow.  We assume a regime $\Rm \gg 1$, in which the magnetic diffusivity has minimal effect on energy-containing length scales over short timescales and the magnetic field is frozen into the fluid.  The induction equation is then
	\begin{equation}
		\pd{\v{B}}{t} = \v{\nabla} \times (\v{v} \times \v{B}) = \v{B} \cdot \v{\nabla} \v{v} - \v{v} \cdot \v{\nabla} \v{B},
	\end{equation}
where the second equality applies for incompressible flow $\v{\nabla} \cdot \v{v} = 0$, and $\v{\nabla} \cdot \v{B} = 0$ has been used.  Then
	\begin{equation}
		\pd{\v{B}}{t} = B_y \pd{U}{y} \unit{x} - U \pd{\v{B}}{x}.
	\end{equation}

Consider the evolution of the magnetic field from some arbitrary initial time $t=0$ over a short increment $\D t$.  The magnetic field at time $\D t$ is
	\begin{align}
		\label{Bperturbation_shorttime}
		\v{B}(\D t) &= \biggl( B_{x} + \D t B_{y} \pd{U}{y} - \D t U \pd{B_{x}}{x} \biggr) \unit{x} \notag \\
				& \quad + \biggl( B_{y} - \D t U \pd{B_{y}}{x} \biggr) \unit{y} + O(\D t^2),
	\end{align}
where the magnetic-field quantities on the RHS are evaluated at $t=0$.  At time $\D t$, we can evaluate the zonally averaged Maxwell stress as
	\begin{align}
		&\ol{B_x(\D t) B_y(\D t)} = \ol{B_{x} B_{y}} + \D t \ol{B_{y}^2} \pd{U}{y} \notag \\
			& \qquad \qquad -\D t \ol{\biggl(U B_{y} \pd{B_{x}}{x} + U B_{x} \pd{B_{y}}{x} \biggr)} + O(\D t^2).
	\end{align}
The third and fourth terms on the RHS combine to become $-\D t U \ol{\partial_x (B_{x} B_{y})}$, which is annihilated by the zonal average.  We obtain 
	\begin{equation}
		\D \ol{B_x B_y} = \D t \ol{B_{y}^2} \pd{U}{y} + O(\D t^2),
		\label{magstress_IVP}
	\end{equation}
where $\D \ol{B_x B_y} \defineas \ol{B_x(\D t) B_y(\D t)} - \ol{B_x(0) B_y(0)}$.  We have thus derived a relation between the Maxwell stress and the shear.  The periodicity constraint is essential in producing this exact result; the physical systems in which zonal flow is known to occur all have such periodicity.  The derivation of \eqnref{magstress_IVP} depends on nothing but the standard MHD induction equation at large $\Rm$.

We would like our theoretical description to involve statistically observable quantities, but \eqnref{magstress_IVP} involves the quantities $\D \ol{B_x B_y}$ and $\ol{B_y^2}$ evaluated at arbitrarily selected points in time, and the increment $\Delta t$, which is not a physically observable quantity associated with the system.  What we do is neglect terms of $O(\D t^2)$ and then replace $\D t$ by the shortest relevant correlation time $\tc$.  The intuition guiding this choice is that the coherent effect of the velocity shear builds up in the magnetic field and in the Maxwell stress until some process stops it.  Various processes can scramble the coherent effect of the shear, including the magnetic diffusion, turbulent inertial effects, and the shearing itself.  In turbulent flow, particularly in hydrodynamic regimes where the magnetic forces are too weak to have significant effect, the turbulent eddy turnover time and perhaps the shearing time are natural correlation times.  Whether the higher-order terms in $\D t$ are negligible in \eqnref{magstress_IVP} depends on the specific regime and the various decorrelation mechanisms.

In light of the preceding paragraph, we make a conceptual switch, from interpreting our result in \eqnref{magstress_IVP} as a literal initial-value calculation to instead viewing it through the lens of time-averaged statistical observables.  With the replacement of $\D t$ by $\tc$, we interpret both the left- and right-hand sides of \eqnref{magstress_IVP} as statistical, time-averaged quantities.  The relation between the shear and the magnetic stress becomes
	\begin{equation}
		\ol{B_x B_y} = \a \ol{B_y^2} \tc \pd{U}{y},
		\label{magstress_shear_relation}	
	\end{equation}
where we have introduced an order-unity constant $\a$.  Hence, the net result is a magnetically originated dynamic viscosity of the form
	\begin{equation}
		\m_m = \a \frac{\ol{B_y^2}}{\m_0} \tc, \label{magneticdynamicviscosity}
	\end{equation}
such that in the zonally averaged momentum equation, the mean shear results in a coherent contribution to the zonally averaged Lorentz force of $\ol{F}_x = \partial_y (\m_m \partial_y U)$.


We stress that the derivation of \eqnref{magstress_IVP} requires no assumptions on the spatial structure of the magnetic field.  The magnetic field could be straight (i.e., a poloidal field), but it could also have arbitrary spatial structure.  $\ol{B_{y}^2}$ incorporates the mean magnetic field as well as all magnetic field perturbations.

\begin{figure}[!t]
		\includegraphics[width=0.5\columnwidth]{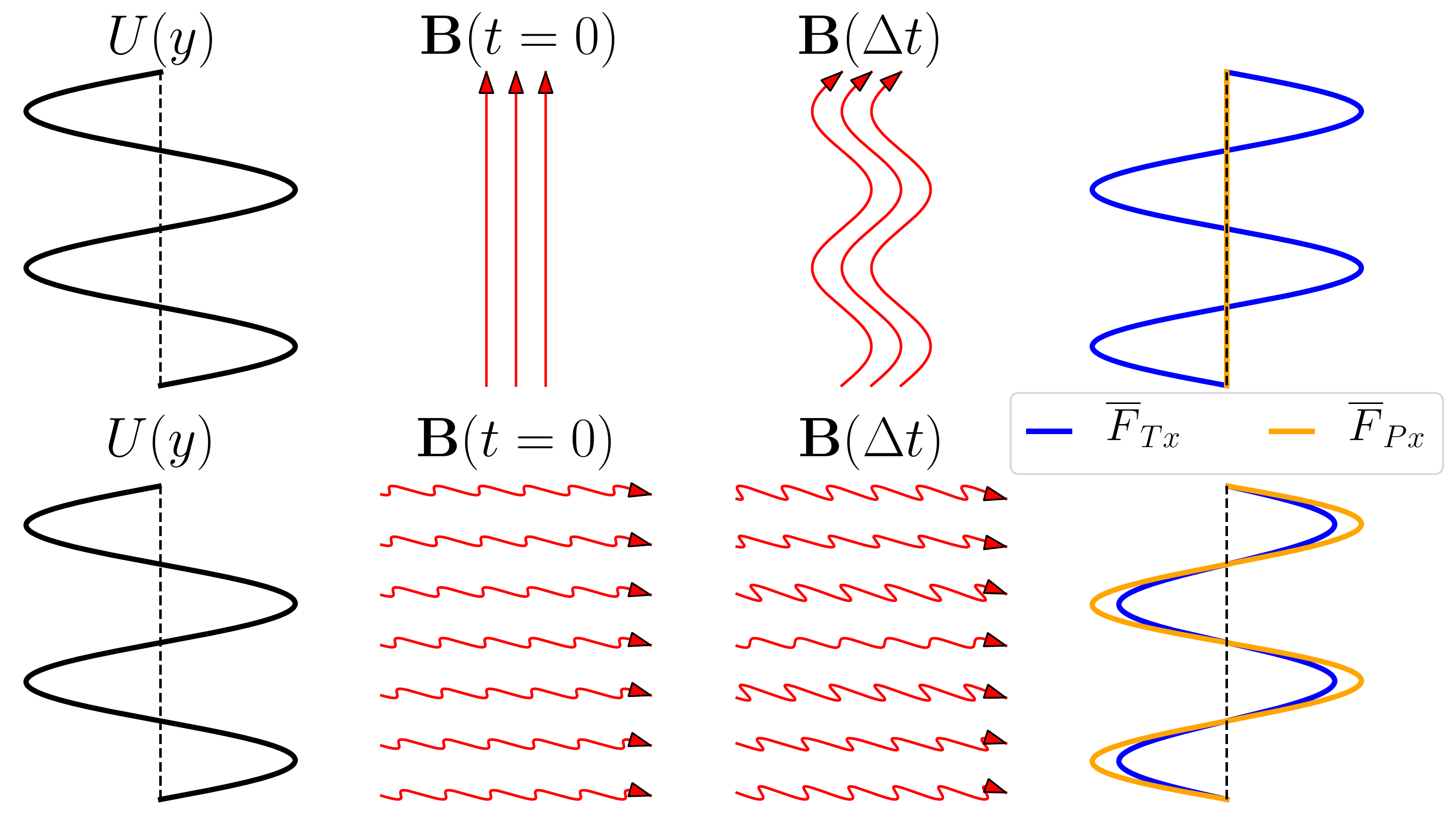}
		\caption{Change in the magnetic field in a short time increment $\D t$ due to a mean shear flow for two different initial configurations, neglecting magnetic diffusivity.  The modification of the field has been exaggerated for visibility.  The resulting $x$ component of the zonally averaged Lorentz force $\ol{F}_x$ in both cases acts as a viscosity to reduce the shear flow.  The decomposition into magnetic tension and pressure forces $\ol{F}_{Tx}$ and $\ol{F}_{Px}$, respectively, is also shown.  The top row shows that from an initially straight field line, $\ol{F}_x$ is due solely to tension (at first order in $\D t$).  The bottom row shows that when the magnetic field is not straight, $\ol{F}_x$ will in general have contributions from both tension and pressure.}
		\label{fig:cartoon}
	\end{figure}

Figure \ref{fig:cartoon} shows the induction of a magnetic field by a shear flow over a short time $\D t$ for two initial field configurations.  Also depicted is the decomposition of the $x$ component of the zonally averaged Lorentz force into the tension and pressure force $\v{J} \times \v{B} = \v{F}_T + \v{F}_P$, where $\v{F}_T = (B^2 / \m_0) \unit{b} \cdot \v{\nabla} \unit{b}$, $\v{F}_P = -\v{\nabla}_\perp B^2 / 2\m_0$, $\unit{b} = \v{B}/B$, and $\v{\nabla}_\perp = (\mathtensor{I} - \unit{b}\unit{b}) \cdot \v{\nabla}$.  When the initial magnetic field lines are straight, $\ol{F}_x$ is due to magnetic tension, akin to the behavior of a shear {\Alfven} wave.  However, in the general case with nonstraight field lines, as might be expected to occur in the tangle of magnetic fields when induction is important and $\Rm \gg 1$, $\ol{F}_x$ includes contributions from both tension and pressure.  In the alternative definition of pressure and tension, where the pressure is isotropic and the tension is given by $\v{B} \cdot \v{\nabla} \v{B} / \m_0$, $\ol{F}_x$ is entirely tension.  Regardless of the initial field configuration, $\ol{F}_x$ still acts as a viscosity, in accordance with~\eqnref{magstress_IVP}.

In the derivation of~\eqnref{magstress_IVP}, we have only imposed the effect of a mean shear flow.  Let us consider what would happen if we included the full fluid velocity field in the induction equation, in which the mean shear flow $U \defineas \ol{\v{v}} \cdot \unit{x}$ is embedded.  In a short time increment, there would be additional contributions to $\D \v{B}$.  Those contributions not correlated across the zonal domain would mostly cancel out when the zonal average of the Maxwell stress is taken.  Hence, it is the mean shear flow that gives rise to a coherent zonally averaged effect.

In Sec.~\ref{sec:simulation_results}, we compare the prediction \eqnref{magstress_shear_relation} for the Maxwell stress with direct numerical simulations.

\section{Negative eddy viscosity}
\label{sec:negative_eddy_viscosity}
For comparison with the magnetic eddy viscosity, it is worthwhile to recall the so-called negative eddy viscosity.  The notion of a hydrodynamic negative eddy viscosity in 2D flow has a long history \cite{starr:1953, kraichnan:1976,gama:1994,krommes:2000}.   A negative eddy viscosity emerges in a variety of mathematical approaches, which can be seen as a reflection of the general tendency for energy to tend towards large scales in 2D \cite{fjortoft:1953}.  The early theoretical literature was primarily concerned with the context of isotropic turbulent flows.  Recent work, however, has revived the negative eddy viscosity to help explain and interpret the physics of the formation and maintenance of zonal flow.

For example, the zonostrophic instability supposes a statistically homogeneous turbulent background and considers the growth of a perturbing zonal flow \cite{farrell:2007, srinivasan:2012, parker:2013, parker:2014, constantinou:2014, fitzgerald:2018}.  For a weak zonal flow of wavenumber $q$ and turbulent background with spectrum $\Phi(\v{k})$, the flow reinforces the perturbation at a rate that schematically looks like $\int d\v{k}\, q^2 (1 - q^2/k^2) g(\v{k}) \Phi(\v{k})$, where $g$ contains anisotropic wave-vector dependence.  At small $q$, this goes like $q^2$.  At larger $q$, there is a cutoff of the form $1 - q^2 / k^2$, which can be traced back to the $\v{v}' \cdot \v{\nabla} \ol{\z}$ term, where $\z$ is the vorticity \cite{bakas:2013-jas,parker:2016}.  Many treatments have neglected this term compared to $\ol{\v{v}} \cdot \v{\nabla} \z'$ under the assumption of long-wavelength mean flows.  We return to discussion of this cutoff at the end of the section.

Manifestations of a negative eddy viscosity are visible at cloud level in both Jupiter and Saturn.  Careful measurements have demonstrated that the Reynolds stress of observed cloud-level velocity fluctuations are positively correlated with the mean shear \cite{salyk:2006,delgenio:2007,delgenio:2012}.  Hence, the turbulent fluctuations appear to transfer energy to the mean flow via Reynolds stress.  Measurements of the turbulent Reynolds stress in laboratory plasma devices lead to similar conclusions~\cite{holland:2006,tynan:2006}.

It is instructive to calculate the Reynolds stress in a similar manner as we did the Maxwell stress.  The calculation also serves as yet another simple demonstration of a negative eddy viscosity.  The details are given in Appendix~\ref{app:negative_eddy_viscosity}.  We obtain a prediction for the Reynolds stress,
	\begin{equation}
		\r \ol{v_x' v_y'} = \g \r \ol{v_y^2} \tc \pd{U}{y},
		\label{reynolds_stress_prediction}
	\end{equation}
where $\g$ is a positive constant of order unity.  This corresponds to a negative dynamic eddy viscosity,
	\begin{equation}
		\m_e = -\g \r \ol{v_y^2} \tc.
		\label{negative_eddy_viscosity}
	\end{equation}

The dynamic eddy viscosity has magnitude approximately equal to the kinetic energy density multiplied by a correlation time.  In pure hydrodynamic flow, the eddy turnover time at the energy-containing scale, $\tt$, is often the appropriate correlation time.  Using $\vt \sim \Lt / \tt$ leads to a mixing-length estimate for the kinematic negative eddy viscosity acting on mean shear flow, $\n_e = -\g \Lt^2 / \tt$. 

Assuming the relevant correlation times in Eqs.~\eqref{magstress_shear_relation} and~\eqref{reynolds_stress_prediction} are equal, the zonally averaged momentum flux due to the turbulent Reynolds and Maxwell stresses is
	\begin{equation}
		\pd{}{t} (\ol{\r v_x}) \bigr|_{\rm turb} = \pd{}{y} \left[ \left( \a \frac{\ol{B_y^2}}{\m_0} - \g \r \ol{v_y^2} \right) \tc \pd{U}{y} \right].
	\end{equation}
Hence, the magnetic eddy viscosity becomes comparable to the negative eddy viscosity when the fluctuating magnetic energy has reached a level roughly similar to that of the fluctuating kinetic energy.  However, in Sec.~\ref{sec:simulation_results} we will see from our 2D simulations that the numerical prefactor $\g$ can be an order of magnitude smaller than $\a$.  Therefore, if the turbulent Reynolds stress is the primary driver of zonal flow, the magnetic viscosity can have a significant influence even when the fluctuating magnetic energy is much smaller.

A refinement of \eqnref{reynolds_stress_prediction} is possible.  The aforementioned $1 - q^2 / k^2$ cutoff in the growth rate motivates a simple modification of \eqnref{reynolds_stress_prediction},
	\begin{equation}
		\r \ol{v_x' v_y'} = \g_2 \r \ol{v_y^2} \tc \hat{L} \left( \pd{U}{y} \right).
		\label{reynolds_stress_prediction_refined}
	\end{equation}
Here, $\hat{L}$ is an operator that multiplies the Fourier amplitude by $1 - q^2 / k_c^2$ for $q \le k_c$, and suppresses the Fourier amplitude for $q > k_c$, where $k_c$ is a characteristic eddy wave number.  This refinement is additionally motivated by the fact that modeling attempts that use only the local shear and neglect higher-order derivatives lead to divergent growth of small scales, but proper inclusion of the cutoff leads to well-behaved models \cite{parker:2016}.

\section{Simulation Setup}
\label{sec:simulation_setup}
We perform a series of numerical simulations to assess our prediction for the Maxwell stress and Reynolds stress. We study the behavior of a 2D incompressible MHD fluid on a magnetized $\b$ plane \cite{tobias:2007}.

The coordinates on the $\b$ plane are $\v{x}\defineas (x,y)$, where $x$ is the azimuthal direction (longitude) and $y$ the meridional direction (latitude).  It is convenient to represent both the velocity and magnetic field by potentials.  The velocity in the rotating frame is written as $\v{v} = \unit{z} \times \v{\nabla} \psi$, where $\psi$ is the stream function.  The vorticity normal to the plane is $\z \defineas \unit{z} \cdot (\v{\nabla} \times \v{v}) = \nabla^2 \psi$.  The magnetic field $\v{B} = \v{B}_0 + \v{\Btilde}$ consists of a constant, uniform background field $\v{B}_0$ and a time-varying component $\v{\Btilde}(\v{x},t)$.  We define a corresponding vector potential $\v{A} = A\unit{z}$ such that $\v{B} = \v{\nabla} \times \v{A}$, where $A = A_0 + \tilde{A}$ and $A_0 = -B_{0y}x + B_{0x}y$.  As we have imposed the presence of a background magnetic field, this 2D system is not subject to antidynamo theorems, and magnetic amplification can result such that $B \gg B_0$.  We perform simulations in which the background magnetic field is either purely toroidal (i.e., aligned in the azimuthal direction, $B_{0y}=0$) or purely poloidal (i.e., aligned in the latitudinal direction, $B_{0x}=0$).

The evolution of the system can be described by a formulation involving vorticity and magnetic potential,
\begin{subequations}
	\label{eq:EOM}
	\begin{gather}
		\pd{\z}{t} + \v{v} \cdot \v{\nabla}(\z + \b y) = -\v{B} \cdot \v{\nabla} \nabla^2 A -\drag\z + \nu \nabla^2 \z + \xi, \label{eq:psi} \\
		\pd{A}{t} + \v{v} \cdot \v{\nabla} A = \eta \nabla^2 A. \label{eq:A} 
	\end{gather}
\end{subequations}
Here, $\b$ is the latitudinal gradient of the Coriolis parameter, $\drag$ is the linear drag, $\nu$ is the viscosity, $\eta$ is the magnetic diffusivity, and $\xi(\v{x},t)$ is a mechanical forcing to excite hydrodynamic fluctuations.  The first term on the RHS of \eqnref{eq:psi} stems from the curl of the Lorentz force.  Equation~\eqref{eq:A} is the magnetic induction equation written for the potential $A$.  For mathematical convenience, we have set the permeability $\mu_0 = 1$ and the mass density $\rho=1$.

We numerically solve \eqnref{eq:EOM} using the code Dedalus~\cite{burns:2019}. We employ periodic boundary conditions in both directions.  During time stepping, we use a high-wave-number spectral filter to remove enstrophy from accumulating at the grid scale.  Unless noted, our domain size is $L_x \times L_y = 4\pi\times4\pi$ at a resolution of $512^2$ gridpoints.  The forcing $\xi$ is statistically homogeneous and isotropic white noise.  It is also small scale, localized near wave vectors with magnitude $k_f = 12$, and injects energy per unit area at a rate $10^{-3}$ (see Ref.~\cite{constantinou:2018} for more details).

The majority of our simulations use nonzero drag $\drag$ on the hydrodynamic flow in order to saturate without filling the largest scales of the box.  However, we also show a case with only viscosity, $\drag=0$, to confirm our results are not sensitive to the presence of drag.

\section{Simulation Results}
\label{sec:simulation_results}

We perform two series of simulations: one with a background toroidal magnetic field $B_0=10^{-2}$, and one with a background poloidal magnetic field $B_0=10^{-3}$.  These runs use $\drag = 10^{-2}$,  $\nu=10^{-4}$, and $\beta = 10$.  For each simulation series, we vary $\eta$.  It is then convenient to define another magnetic Reynolds number that is referenced to the hydrodynamic limit,
	\begin{equation}
		\Rm_0 \defineas \frac{L_0 V_0}{\eta},
	\end{equation}
where $L_0$ and $V_0$ are a characteristic length scale and velocity scale in a regime where the magnetic field has no influence, i.e., large $\eta$.  For $L_0$ we use a typical eddy length scale computed from the eddy energy spectrum as $\Lt \defineas [ \int d\v{k}\, k^2 E(\v{k}) / \int d\v{k}\, E(\v{k})]^{-1/2}$.
For $V_0$ we use the rms eddy velocity $\vt$.  
Our simulations yield $\Lt = 0.14$ and $\vt = 0.21$, giving $\Rm_0 = 0.03 / \eta$.  Our reported values of $\mathcal{A}$ are computed as the total magnetic energy divided by the total kinetic energy.

As we decrease $\eta$ from large values, we move through the three regimes depicted in Figure~\ref{fig:parameterspace}.  First, is the regime of large $\eta$, $\Rm_0 < 1$, and magnetic energy much smaller than kinetic energy, $\mathcal{A} \ll 1$.  As $\eta$ is reduced, a second regime is entered, where $\Rm_0 > 1$, and still $\mathcal{A} \ll 1$.  Then, as $\eta$ is decreased even further, $\Rm_0 \gg 1$ and $\mathcal{A} \approx 1$.  The intermediate regime is the one where the assumptions behind the magnetic eddy viscosity are valid.

\begin{figure}
		\centering
		\includegraphics[width=0.85\columnwidth]{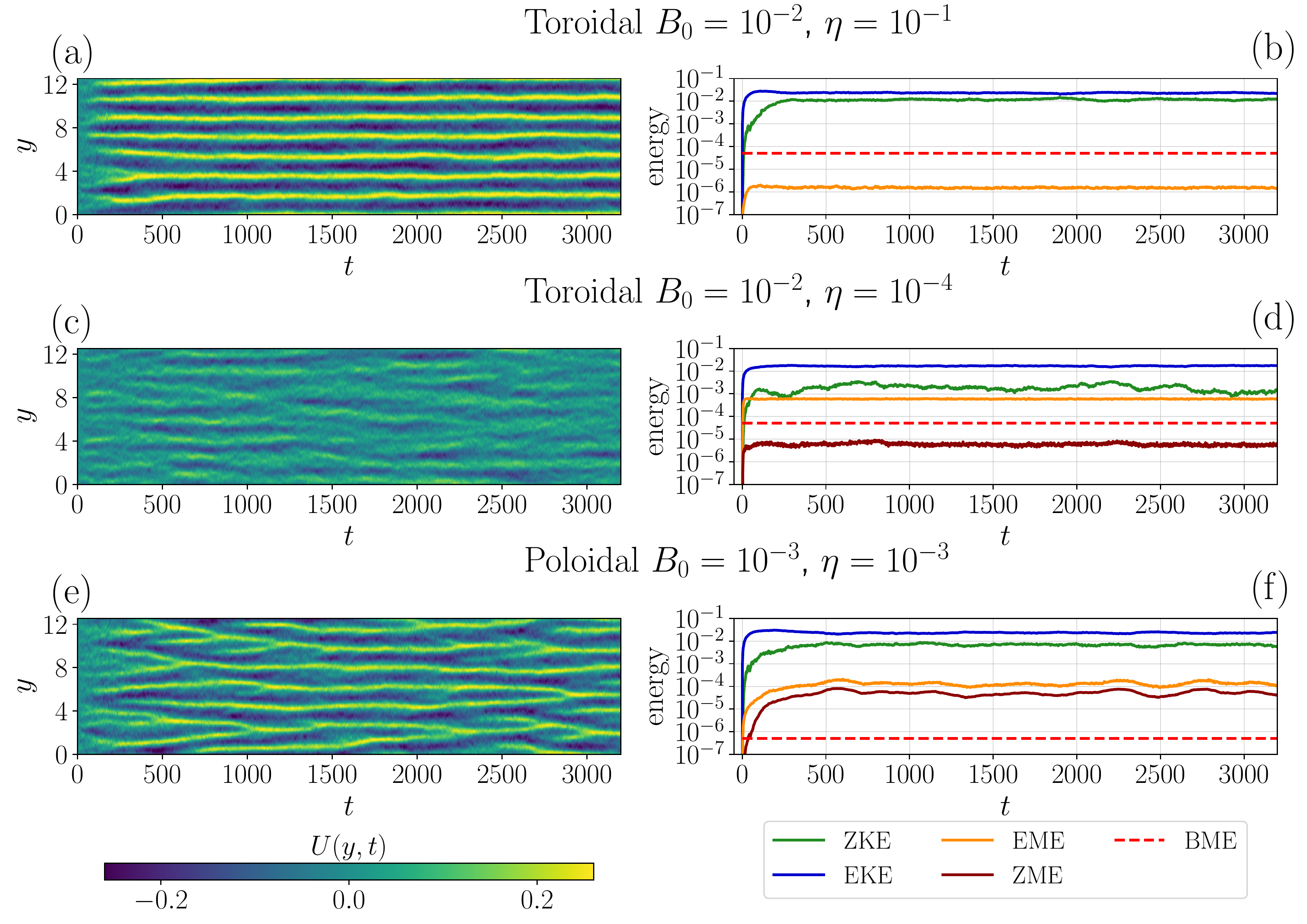}
		\caption{Results from three representative simulations.  The left column shows space-time plots of the zonal flow $U$ and the right column shows time traces of averaged energy densities associated with the zonal flow (ZKE) and the hydrodynamic eddies (EKE), the background magnetic field (BME), the zonally averaged magnetic field (ZME), and the eddy magnetic field (EME).  (a) and (b) Case at $\Rm_0=0.3$ in which magnetic fluctuations are too weak to have an effect on the hydrodynamic flow.  (c) and (d) Effect of magnetic field on the zonal flow with toroidal background field and $\Rm_0=300$.  (e) and (f) Case with poloidal background field and $\Rm_0 = 30$.}
		\label{fig:representativecases-energy}
	\end{figure}

Figure~\ref{fig:representativecases-energy} shows results from three representative simulations.  The left column shows space-time diagrams of the zonal flow $U(y,t) = \ol{v}_x$.  The right column shows time series of the different contributions to the domain-averaged energy density: zonal-flow kinetic energy (ZKE) $\langle U^2 / 2 \rangle$, eddy kinetic energy (EKE) $\langle (v_x'^2 + v_y'^2) / 2 \rangle$, zonal magnetic energy (ZME) $\bigl \langle \ol{\Btilde}_x^2 / 2 \bigr \rangle$, eddy magnetic energy (EME) $\langle (B_x'^2 + B_y'^2) / 2 \rangle$, and the background magnetic energy (BME) $B_0^2 / 2$.  Here, $\langle f \rangle \defineas \int d\v{x}\, f / L_x L_y$ denotes a domain average of quantity $f$.  

Figures \ref{fig:representativecases-energy}(a) and \ref{fig:representativecases-energy}(b) show a case with $\Rm_0=0.3$. Figure~\ref{fig:representativecases-energy}(a) shows that zonal jets form spontaneously.  After an initial transient time of about $5/\drag = 500$, seven jets equilibrate at finite amplitude and remain coherent throughout the rest of the simulation.  The eddy magnetic energy is about 4 orders of magnitude smaller than the kinetic energy of the flow [Fig.~\ref{fig:representativecases-energy}(b)].  Figures \ref{fig:representativecases-energy}(c) and \ref{fig:representativecases-energy}(d) show a case with toroidal background field at $\Rm_0= 300$.  The magnetic energy is somewhat smaller than the kinetic energy, $\mathcal{A} = 0.05$.  The zonal flow is mostly suppressed; it has smaller amplitude and is less coherent.  Figures \ref{fig:representativecases-energy}(3) and \ref{fig:representativecases-energy}(f) show a case with poloidal background field at $\Rm_0 = 30$. Here, the zonal kinetic energy is not reduced from the hydrodynamic case as in Fig.~\ref{fig:representativecases-energy}(a), but the coherence of the zonal flow is clearly reduced in Fig.~\ref{fig:representativecases-energy}(e).  There is a continual cycle of jet merging and formation of new jets, and the zonal flow never settles down to a steady state.  Interestingly, this behavior occurs even when $\mathcal{A} = 0.006$.

\subsection{Maxwell stress}
We compare the Maxwell stress predicted by \eqnref{magstress_shear_relation} with the actual stress observed in the simulations.  To facilitate the comparison, we perform a least-squares best fit for the single free parameter $\a$, which serves merely as an overall constant of proportionality and does not affect the spatial structure.  The fit is performed on the Maxwell stress itself, not its derivative.

\begin{figure}
		\includegraphics[width=0.8\columnwidth]{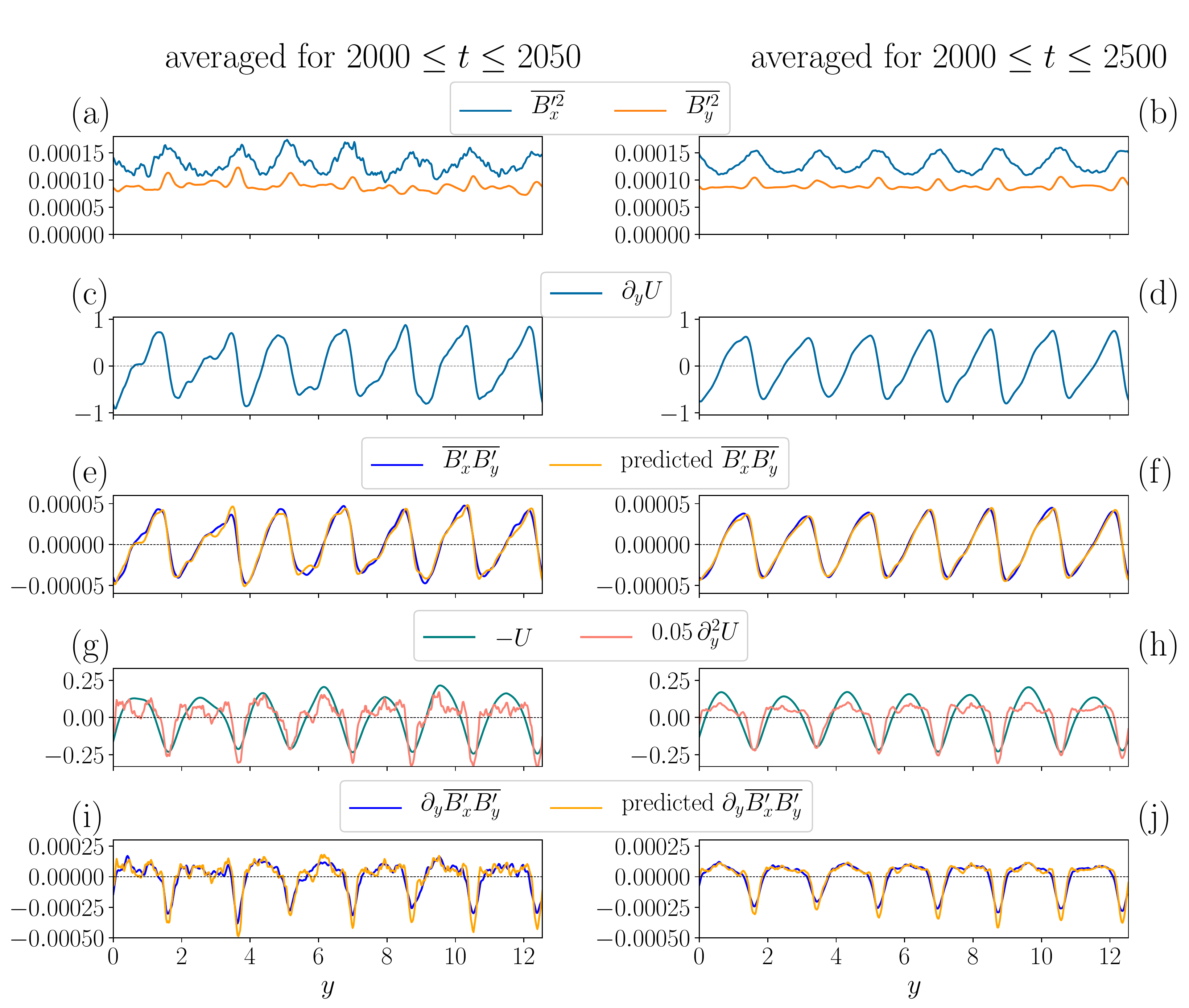}
		\caption{Results for the simulation with toroidal $B_0 = 10^{-2}$, $\eta = 10^{-3}$ ($\Rm_0 = 30$), and $\mathcal{A} = 0.005$.  The left and right columns show short-time and long-time averages, respectively.  (a) and (b) Zonally averaged magnetic eddy energy components $\ol{B_x'^2}$ and $\ol{B_y'^2}$.  (c) and (d) Zonal flow shear $\partial_y U$.  (e) and (f) Maxwell stress $\ol{B_x'B_y'}$ and the prediction of \eqnref{magstress_shear_relation}.  (g) and (h) Structure of $-U$ and $\partial_y^2U$.  (i) and (j) Divergence of the Maxwell stress $\partial_y\ol{B_x'B_y'}$ and the derivative of the prediction of \eqnref{magstress_shear_relation}.   The proportionality constant $\a$ is $1.02$ for the short-time average and $1.04$ for the long-time average.}
		\label{fig:representativecases-magneticstress-toroidal}
	\end{figure}

Figure~\ref{fig:representativecases-magneticstress-toroidal} examines a simulation with toroidal background field.  For this simulation, $\Rm_0 = 30$ and $\mathcal{A} = 0.005$.  The curves labeled ``predicted $\ol{B_x' B_y'}$'' are computed using correlation time $\tc = \Lt / \vt$.  Both short-duration averages (50 time units, left column) and long-duration averages (500 time units, right column) are shown for this comparison to illustrate that there is no substantial difference. The proportionality constant $\a = 1.02$ for the short-time average and 1.04 for the long-time average.  The predictions for the Maxwell stress from \eqnref{magstress_shear_relation} are in very good agreement with those from the simulations.  $\ol{B_y'^2}$ does not vary much with~$y$ [Figs.~\ref{fig:representativecases-magneticstress-toroidal}(a) and \ref{fig:representativecases-magneticstress-toroidal}(b)], and therefore the Maxwell stress closely resembles the mean flow shear; compare Figs.~\ref{fig:representativecases-magneticstress-toroidal}(c) and \ref{fig:representativecases-magneticstress-toroidal}(d) with Figs.~\ref{fig:representativecases-magneticstress-toroidal}(e) and \ref{fig:representativecases-magneticstress-toroidal}(f).  Additionally, by comparing the spatial dependence of $\partial_y \ol{B_x' B_y'}$ [Figs.~\ref{fig:representativecases-magneticstress-toroidal}(i) and \ref{fig:representativecases-magneticstress-toroidal}(j)] with $-U$ and $\partial_y^2 U$ [Figs.~\ref{fig:representativecases-magneticstress-toroidal}(g) and \ref{fig:representativecases-magneticstress-toroidal}(h)], we see that $\partial_y\ol{B'_xB'_y}$ more closely resembles $\partial_y^2 U$.  Thus, the Lorentz force acts like viscosity rather than like drag.

\begin{figure}
	 	\includegraphics[width=0.8\columnwidth]{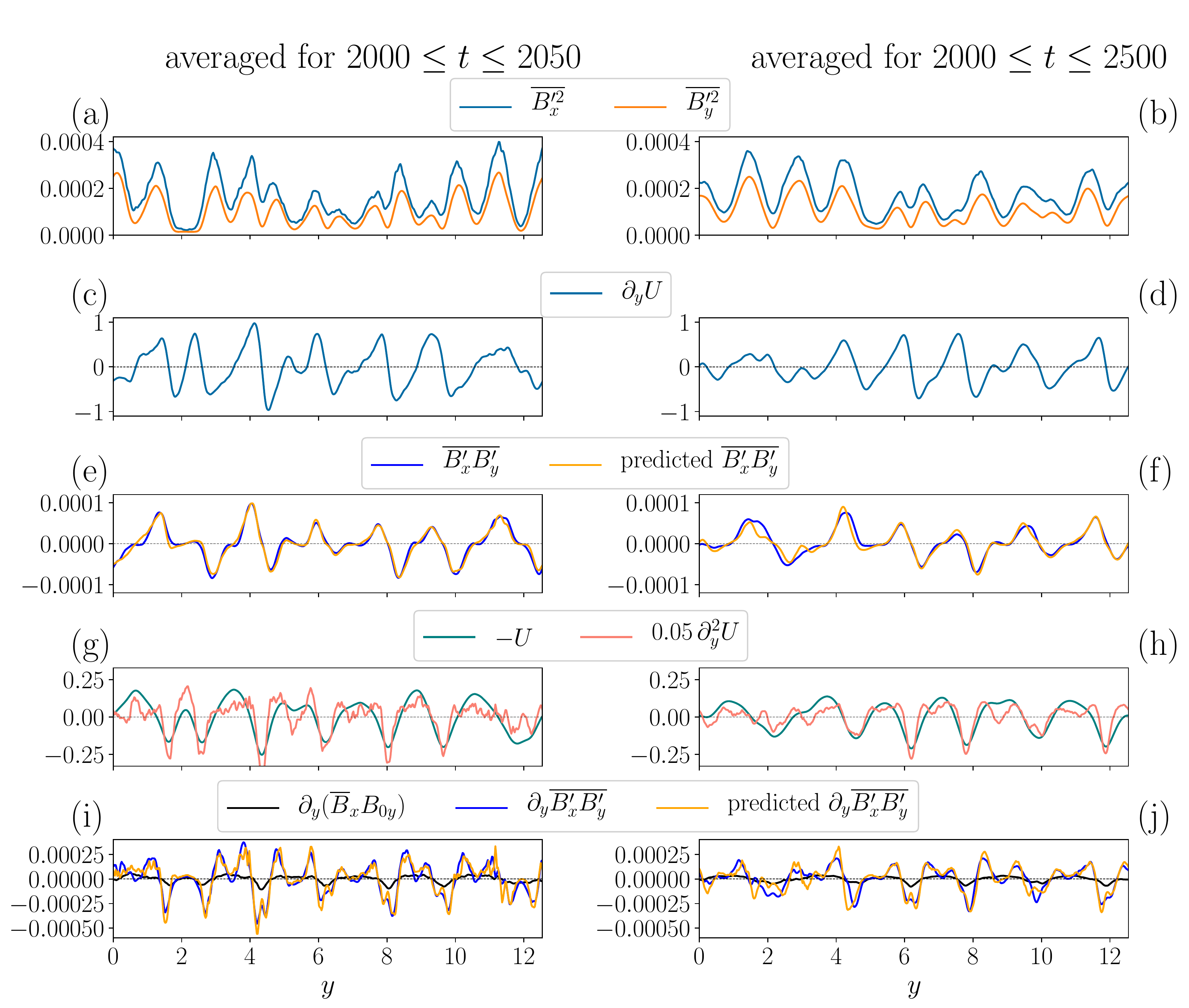}
		\caption{Same as Fig.~\ref{fig:representativecases-magneticstress-toroidal}, for the simulation with poloidal $B_0 = 10^{-3}$, $\eta = 10^{-3}$ ($\Rm_0 = 30$), and $\mathcal{A}=0.006$.  In (i) and (j), the term $\partial_y(\ol{B}_xB_{0y})$ is also shown.  The proportionality constant $\a$ is $1.05$ and  $1.17$ for the short- and long-time averages, respectively.}
		\label{fig:representativecases-magneticstress-poloidal}
	\end{figure}

Figure~\ref{fig:representativecases-magneticstress-poloidal} is much the same, but for a poloidal background field, with $\Rm_0 = 30$ and $\mathcal{A} = 0.006$.  There is an additional subtlety here regarding the calculation of the Maxwell stress.  The prediction in Sec.~\ref{sec:magneticviscosity} uses the total magnetic field $\v{B}$.  For a poloidal background field, $\ol{B_x B_y} = \ol{B}_x B_{0y} + \ol{B_x' B_y'}$ (whereas for a toroidal background field, $\ol{B_x B_y} = \ol{B_x' B_y'}$).  However, \eqnref{magstress_shear_relation} can be used as a prediction for $\ol{B_x' B_y'}$ if we replace $\ol{B_y^2}$ with $\ol{B_y'^2}$.  The figure shows the prediction for $\ol{B_x' B_y'}$ rather than $\ol{B_x B_y}$, though this distinction has little quantitative significance here because $B_0 \ll \Btilde$ for $\Rm \gg 1$.

In Fig.~\ref{fig:representativecases-magneticstress-poloidal}, the prediction for the Maxwell stress agrees quite well with the actual values.  The proportionality constant $\a = 1.05$ for the short-time average and 1.17 for the long-time average.  It is interesting to note that unlike the toroidal case, $\ol{B_y'^2}$ has significant variation in $y$.  This dependence leaves a noticeable imprint in the Maxwell stress, showing that $\ol{B_x' B_y'}$ is not merely proportional to the flow shear $\partial_y U$.  Figures~\ref{fig:representativecases-magneticstress-poloidal}(i) and \ref{fig:representativecases-magneticstress-poloidal}(j) show, in addition to the derivative of the fluctuating stresses, the term $\partial_y(\ol{B}_xB_{0y})$.  This term, which is responsible for magnetic drag at $\Rm \ll 1$, is much smaller than $\partial_y\ol{B'_x B'_y}$ here.


\begin{figure}[!t]
	  \centering
		\includegraphics[width=0.85\columnwidth]{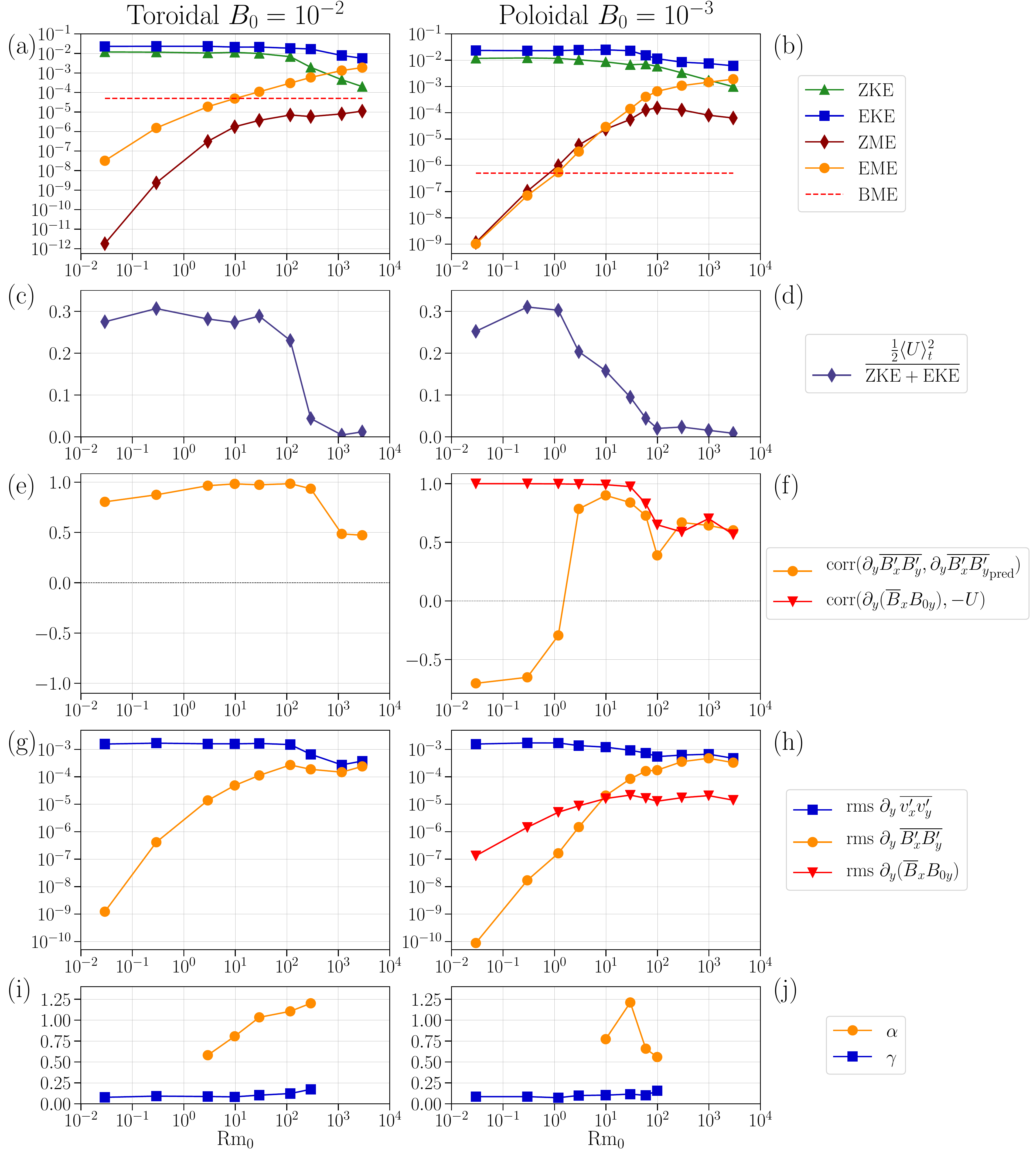}
		\caption{Results for several simulations with varying $\eta$ ($\Rm_0 = 0.03/\eta$).  The left column shows the case with background toroidal field and the right column with poloidal field.  (a) and (b) Steady-state energies (see the text for definitions of the various terms).  $\mathcal{A}$ is small for $\Rm_0 \lesssim 300$.  (c) and (d) Fraction of kinetic energy in a coherent zonal flow, where $\langle \cdot \rangle_t$ is a time average.  (e) and (f) Correlations of the magnetic stresses observed in the simulations with the prediction of \eqnref{magstress_shear_relation}.  For the case of a poloidal field, the correlation of $\partial_y (\ol{B}_x B_{0y})$ with $-U$ is 1 for $\Rm_0 < 1$; this is magnetic drag.  (g) and (h) The rms (over $y$) of the relevant stresses, showing that the Maxwell stress rises to be sufficiently strong to counteract the Reynolds stress at large enough $\Rm$.  (i) and (j) Best-fit proportionality constants $\a$ and $\g$, where $\a$ is shown only for the cases in the magnetic eddy viscosity regime: $\Rm_0 > 1$, but small enough such that a coherent mean zonal flow still exists.  All quantities were averaged over  $2000\le t\le 3000$.} 
		\label{fig:results_vs_eta}
	\end{figure}

Figure~\ref{fig:results_vs_eta} shows simulations results at different values of $\Rm_0$.  The left column shows results in the statistically steady state for a toroidal background field.  The regime of magnetic eddy viscosity is valid for $3 \lesssim \Rm_0 \lesssim 300$.  The upper bound is determined where the magnetic field has a significant influence on the flow and the condition $\mathcal{A} \ll 1$ no longer holds.  Figure~\ref{fig:results_vs_eta}(e) shows the correlation coefficient between the derivative of the Maxwell stress and the prediction, which is close to 1 in the magnetic eddy viscosity regime.  Figure~\ref{fig:results_vs_eta}(g) shows the rms values of the Reynolds stress and Maxwell stress, which exhibit a similar dependence as the kinetic energy and magnetic energy. The proportionality constant $\a$ for both the toroidal and poloidal cases is approximately equal to 1, as seen in Figs.~\ref{fig:results_vs_eta}(i) and~\ref{fig:results_vs_eta}(j).

The right column depicts the case with a poloidal background field.  The details are similar, with the regime of magnetic eddy viscosity existing for $10 \lesssim \Rm_0 \lesssim 100$.  Figure~\ref{fig:results_vs_eta} also demonstrates that the Lorentz force transitions from a magnetic drag at low $\Rm_0$ to magnetic viscosity at high $\Rm_0$.  Figure~\ref{fig:results_vs_eta}(f) shows that at low $\Rm_0$, the correlation of $\partial_y (\ol{B}_x B_{0y})$ with $-U$ is equal to 1, indicating drag.  A straightforward calculation shows the amplitude of the magnetic drag at low $\Rm_0$ is expected to be $\ol{F}_x = -(B_0^2 / \eta \m_0) U$ (momentarily returning to dimensional quantities); this amplitude is quantitatively recovered in the simulation.  As $\Rm_0$ increases, the Lorentz force due to fluctuating magnetic fields, $\partial_y \ol{B_x' B_y'}$, correlates well with the predicted magnetic viscosity.  When $\Rm_0 \gtrsim 10$, the viscouslike fluctuating force begins to dominate, as seen in Fig.~\ref{fig:results_vs_eta}(h).

Even though the kinetic energy is much greater than the magnetic energy for $\Rm_0 < 300$, Fig.~\ref{fig:results_vs_eta}(d) shows how a poloidal background field affects the coherence of the zonal flow even starting at $\Rm_0 = 3$.  Figure~\ref{fig:results_vs_eta}(d) is computed using the time average $\langle U \rangle_t^2/2$, which is reduced compared to the zonal kinetic energy when the spatial structure of the zonal flow is not steady in time [see Figs.~\ref{fig:representativecases-energy}(a) and \ref{fig:representativecases-energy}(e)].  The longer the duration of the time average, the more $\langle U \rangle_t^2$  will be reduced by the decoherence.

\begin{figure}
		\includegraphics[width=0.85\columnwidth]{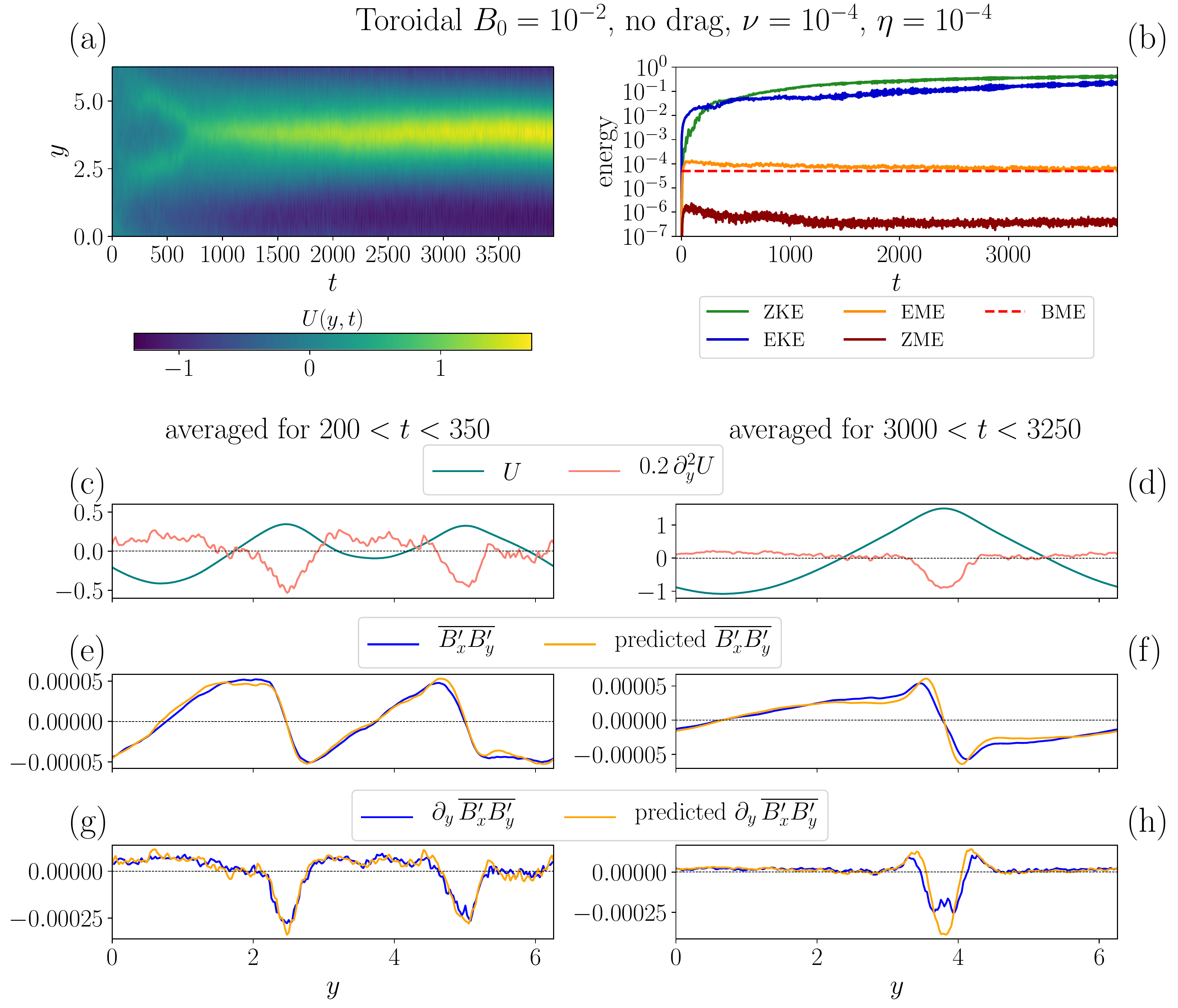}
		\caption{Simulation without frictional drag, with toroidal $B_0 = 10^{-2}$ and parameters $\nu = 10^{-4}$ and $\eta = 10^{-4}$.  In this simulation, $\Lt = 0.56$, $\vt=0.46$, $\Rm \approx 2600$, and $\mathcal{A} \approx 6 \times 10^{-4}$.  See the text for more details.}
		\label{fig:nodrag}
	\end{figure}

To check whether the prediction for the Maxwell stress is sensitive to the presence of frictional drag, we also perform a simulation with $\kappa = 0$ but maintaining nonzero viscosity.  The results are shown in Fig.~\ref{fig:nodrag}.  This simulation uses domain size $L_x \times L_y = 2\pi \times 2\pi$, numerical resolution of $256^2$, and parameter values $\b = 2$, $\n = 10^{-4}$, and $\eta = 10^{-4}$.  The background magnetic field is toroidal, with $B_0 = 10^{-2}$.  Without drag, the structures tend to fill the box size, and only a single jet exists at late times.  Time-averaging around $t=2500$, we find $\Lt = 0.56$, $\vt=0.46$, and hence $\Rm \approx 2600$.  The magnetic energy is much smaller than the kinetic energy, with $\mathcal{A} \approx 6 \times 10^{-4}$.

Equation~\eqref{magstress_shear_relation} accurately predicts the Maxwell stress in this simulation as well.  This is true both at early times $200 < t < 350$, when there are two jets, and at late times, $3000 < t < 3250$, after the two jets have merged so that a single wavelength fills the domain.  Moreover, at late times, the spatial structure of $U$ and $\partial_y^2 U$ are significantly different, and $\partial_y \ol{B_x' B_y'}$ is clearly closer to $\partial_y^2 U$.  We also observe that \eqnref{magstress_shear_relation} holds even though the kinetic energy has not yet saturated under viscous dissipation, because the timescale setting the magnetic quasi-equilibrium is much shorter than the viscous time.	
	
\subsection{Reynolds stress}
\begin{figure}[t]
		\includegraphics[width=0.6\columnwidth]{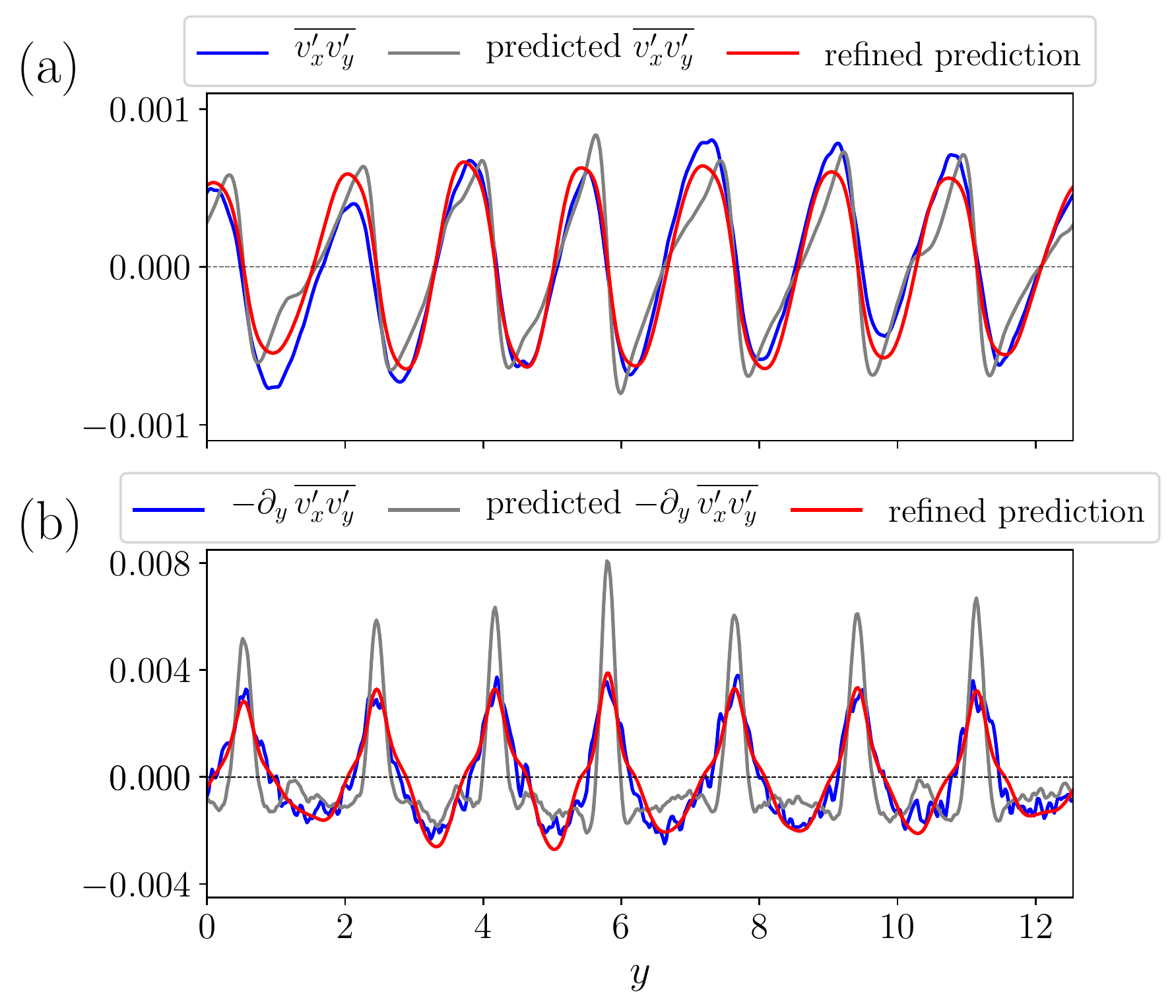}
		\caption{Comparison of the zonally averaged Reynolds stress in a simulation with the prediction of \eqnref{reynolds_stress_prediction} and the refined prediction of \eqnref{reynolds_stress_prediction_refined}.  Shown are (a) the Reynolds stress and (b) its divergence.  This simulation used a toroidal background magnetic field $B_0=10^{-2}$ and $\eta=10^{-2}$ ($\Rm_0=3$). The proportionality constants are $\g=0.08$ and $\g_2=0.13$.  Quantities were averaged over $2000\le t\le 2500$.}
		\label{fig:representativecases-reynoldsstress}
	\end{figure}

We also examine the negative eddy viscosity in our simulations to determine whether the relation in \eqnref{reynolds_stress_prediction} holds in our simulations.  Figure~\ref{fig:representativecases-reynoldsstress} shows results for a simulation with a background toroidal field of $B_0 = 10^{-2}$.  The system has $\mathcal{A} \ll 1$ and so the magnetic field has little influence on the flow.  Correlations induced in the fluctuating Reynolds stress by the zonal flow are picked out by zonal averaging.  The zonally averaged Reynolds stress and its divergence are in overall good agreement with the predictions from \eqnref{reynolds_stress_prediction}, with a best-fit $\g = 0.08$.  Observe that the proportionality constant for the Reynolds stress $\g$ is about 10 times smaller than the proportionality constant for the Maxwell stress~$\a$ [see Figs.~\ref{fig:results_vs_eta}(i) and \ref{fig:results_vs_eta}(j)].

Equation \eqref{reynolds_stress_prediction} produces a noticeable over-prediction of the jet forcing $-\partial_y \ol{v_x' v_y'}$ at the eastward peaks of the jets.  Figure~\ref{fig:representativecases-reynoldsstress} also shows the refined prediction of \eqnref{reynolds_stress_prediction_refined}.  The refined prediction removes the over-estimate of $-\partial_y \ol{v_x' v_y'}$ at the eastward jets and shows much better agreement with the actual value.  We have used $k_c = 1/\Lt = 7.1$, so the only free parameter is the overall proportionality factor which, for this case, is $\g_2 = 0.13$.

\section{Generalization to three dimensions}
\label{sec:3d_generalization}
The simple derivation of magnetic eddy viscosity in Sec.~\ref{sec:magneticviscosity} assumed a 2D Cartesian system.  Here, we follow a similar procedure to generalize \eqnref{magstress_IVP} to three dimensions in Cartesian, cylindrical, and spherical coordinates.  We find the relevant components of the Maxwell stress tensor that are required to compute the zonally averaged Lorentz force.  As in Sec.~\ref{sec:magneticviscosity}, the following equations are exact to first order in $\D t$ and depend only on the large-$\Rm$ limit of the MHD induction equation.

In Cartesian coordinates, let $\v{v} = U(y, z) \unit{x}$.  A zonal average is over $x$, and periodicity in $x$ assumed.  We obtain
\begin{subequations}
	\label{3dgeneralization_cartesian}
	\begin{align}
		\D \ol{B_x B_y} &= \D t \biggl[ \ol{B_{y}^2} \pd{U}{y} + \ol{B_{y}B_{z}} \pd{U}{z} \biggr] + O(\D t^2), \\
		\D \ol{B_x B_z} &= \D t\biggl[ \ol{B_{z}^2} \pd{U}{z} + \ol{B_{y}B_{z}} \pd{U}{y} \biggr] + O(\D t^2).
	\end{align}
\end{subequations}
In cylindrical coordinates $(s, \p, z)$, let $\v{v} = U(s, z) \bm{\hat{\p}}$, and a zonal average is over the azimuthal angle $\p$.  We find
\begin{subequations}
	\label{3dgeneralization_cylindrical}
	\begin{align}
		\D \ol{B_s B_\p} &= \D t \left[ s \ol{B_s^2} \pd{}{s}\left(\frac{U}{s} \right) + \ol{B_s B_z} \pd{U}{z}  \right] + O(\D t^2), \\
		\D \ol{B_z B_\p} &= \D t \left[\ol{B_z^2} \pd{U}{z} + s\ol{B_s B_z} \pd{}{s} \left( \frac{U}{s} \right)  \right] + O(\D t^2).
	\end{align}
\end{subequations}
In spherical coordinates ($r, \th, \p)$, let $\v{v} = U(r, \th) \bm{\hat{\p}}$, and again a zonal average is over $\p$.  We obtain
\begin{subequations}
		\label{3dgeneralization_spherical}
	\begin{widetext}
	\begin{align}
		\D \ol{B_r B_\p} &= \frac{\D t}{r} \biggl[ r^2 \ol{B_r^2} \pd{}{r}\left(\frac{U}{r}\right) + \ol{B_r B_\th} \sin\th \pd{}{\th}\left( \frac{U}{\sin\th}\right) \biggr] + O(\D t^2), \\
		\D \ol{B_\th B_\p} &= \frac{\D t}{r} \biggr[ \ol{B_\th^2} \sin\th \pd{}{\th}\left( \frac{U}{\sin\th}\right) + r^2 \ol{B_r B_\th} \pd{}{r}\left(\frac{U}{r}\right)  \biggr] + O(\D t^2).
	\end{align}
	\end{widetext}
\end{subequations}
	
We make several observations.  First, in cylindrical or spherical geometry, the first-order coherent effect in the Maxwell stress due to $U$ vanishes if $U \sim s \sim r\sin\th$, i.e., solid body rotation.  Second, one can straightforwardly compute that in the cross terms $\ol{B_y B_z}$, $\ol{B_s B_z}$, or $\ol{B_r B_\th}$, there is no first-order coherent effect due to~$U$.  One might expect that the fluctuating field's contribution to this term averages to approximately zero.  A mean magnetic field could lead to nonzero correlation such as $\ol{B_r B_\th}$, although in a regime $\Rm \gg 1$ the $\ol{B_r^2}$ and $\ol{B_\th^2}$ terms likely dominate.  Third, the computation of the zonally averaged Lorentz force in non-Cartesian geometries, $\ol{F}_\p = \ol{\v{\nabla} \cdot \mathtensor{T} \cdot \bm{\hat{\p}}}$, introduces additional geometric factors in the tensor divergence.  Fourth, for rotation about an axis, $U = s \W$, Eqs.~\eqref{3dgeneralization_cylindrical} and \eqref{3dgeneralization_spherical} may be written compactly as
	\begin{equation}
		\D \ol{\v{B}_p \cdot \unit{t} B_\p} = s \D t \ol{\v{B}_p \cdot \unit{t} \v{B}_p \cdot \v{\nabla} \W} + O(\D t^2),
		\label{3dgeneralization_vector}
	\end{equation}
where $\v{B}_p$ is the poloidal (nonazimuthal) part of the magnetic field and $\unit{t}$ is a unit vector in the plane orthogonal to $\unit{\p}$, e.g., $\unit{t} = \unit{s}, \unit{z}, \unit{r}, \bm{\hat{\th}}$.

Equation \eqref{3dgeneralization_vector} functions as a dynamical generalization of Ferraro's law of isorotation, which states that for an axisymmetric magnetic field and axisymmetric azimuthal flow with angular velocity $\W (s, z)$, a steady state is possible only when $\v{B}_p \cdot \v{\nabla} \W = 0$~\cite{ferraro:1937}.  Hence, Ferraro's law has been cited for the notion that magnetic fields tend to suppress differential rotation (although this is not required; the law merely requires that $\v{B}_p$ lies on surfaces of constant $\W$).  Our derivation of a magnetic eddy viscosity generalizes the isorotation law because it (i) allows for turbulent dynamics, (ii) does not assume axisymmetry of the magnetic field, and (iii) provides an estimate for the magnitude of the resulting force that counteracts differential rotation.  The isorotation law is recovered from \eqnref{3dgeneralization_vector} if $\v{B}_p$ is axisymmetric.

Under appropriate physical conditions where second-order effects can be neglected, one can apply the same arguments as in Sec.~\ref{sec:magneticviscosity} to relate the coherent part of the statistically averaged Maxwell stresses to the mean shear flow.  If the first-order truncation is applicable, then it is clear that a differentially rotating body will induce coherent Maxwell stresses that act to redistribute angular momentum in a way similar to a viscosity.

While the prediction of the Maxwell stress was verified in 2D MHD $\b$-plane simulations in Sec.~\ref{sec:simulation_results}, comparing the equations of this section with 3D numerical simulations is out of scope of the present work.  The expressions in Eqs.~\eqref{3dgeneralization_cartesian}--\eqref{3dgeneralization_spherical} could be directly compared with the Maxwell stress observed in 3D simulations to determine the applicability of the magnetic eddy viscosity in a particular regime.


\section{Estimates for Jupiter and Saturn}
\label{sec:planet}
Could the magnetic eddy viscosity regime apply in some planetary interiors and perhaps be responsible for suppression of differential rotation?  We consider this question along with some Jupiter and Saturn data.  We explore the possibility that $\Rm \gg 1$ occurs in regions with significant mean flow (implicitly with $\mathcal{A} \ll 1$).  This analysis is restricted to the upper atmosphere as the deeper planetary dynamo region likely has $\mathcal{A} \sim O(1)$.

There is currently a great deal of uncertainty about the dynamics below the cloud-top level in these two gas giants.  While the flow speed is expected to decrease with depth due to the increasing mass density, for this estimate we will use the observed outer-atmosphere values.

Cloud tracking of Jupiter's and Saturn's upper-level atmosphere showed that eddies act to reinforce the zonal jets with an effective eddy viscosity $\nu_{e,J} \approx -10^6\,\rm m^2\,s^{-1}$ for Jupiter \cite{salyk:2006} and $\nu_{e,S} \approx -2\times 10^5\,\rm m^2\,s^{-1}$ for Saturn~\cite{delgenio:2007,delgenio:2012}.  We observe that $\n_{e,J}$ is roughly equal to the mixing-length estimate $\n_e = -\g \Lt^2 / \tt$, using $\Lt = 1000$ km and $\tt = 2.5 \times 10^5$ s observed at the Jupiter cloud tops.  Supposing the negative eddy viscosity is the sole driving force of the zonal flow, we ask: if the negative eddy viscosity is to be counterbalanced solely by magnetic eddy viscosity, then (i) what is the strength of the magnetic field needed to do so and (ii) how deep below the cloud tops of the two gas giants do we expect such strong magnetic fields?
 
The mass density increases rapidly beneath the cloud tops. We use a number intermediate within the expected density range for both Jupiter and Saturn, $\r \approx 200$ kg\,m$^{-3}$ \cite{french:2012,anderson:2007,mankovich:2019}.  For the correlation time, we use the turbulent eddy turnover time $\tt = 2.5 \times 10^5$ s.  This time is not that different from the typical shear time associated with the surface-level zonal jets $t_{\textrm{shear}, J}\approx 2\times 10^5\,\rm s$ and $t_{\textrm{shear}, S}\approx 5\times10^4\,\rm s$~\cite{vasavada:2005,delgenio:2007}.  We also use typical mean magnetic field values $B_{0,J}=4\times 10^{-3}\,\rm T$~\cite{moore:2018} and $B_{0,S}=2\times 10^{-5}\,\rm T$~\cite{gombosi:2009}. With these in hand, the kinematic magnetic eddy viscosity $\n_m = \m_m / \r$ computed with the mean magnetic field values is 2--4 orders of magnitude smaller than the corresponding negative eddy viscosity. Thus, magnetic viscosity can only start becoming important if we take into account the magnetic fluctuations.

For convenience, we assume the scaling $B^2 \sim B_0^2 \text{Rm}$ at $\Rm \gg 1$.  Accounting for the factor of $\Rm$, we estimate for Jupiter that $\nu_m\approx\nu_e$ requires $\Rm_J\approx 60$ and for Saturn requires $\Rm_S\approx 700$.  Hence these results are consistent with the assumption $\Rm \gg 1$.  Then, using the outer-atmosphere typical eddy length scale $\Lt$ and speed $\vt$, we compute that the required $\Rm$ values correspond to magnetic diffusivities of $\eta_J \approx 6\times10^4\,\rm m^2\,s^{-1}$ and $\eta_S \approx 6\times10^3\,\rm m^2\,s^{-1}$.   Finally, from Refs.~\cite{french:2012,liu:2008,cao:2017} we find that such magnetic diffusivity values are achieved at approximately $0.96R_J$ and $0.8 R_S$. Therefore, under the assumptions we have made, we expect zonal jets would be magnetically terminated by these depths.  These estimates are not inconsistent with the recent gravitometric estimates of depths of $0.96 R_J$ and $0.85 R_S$ \cite{kaspi:2018,guillot:2018,galanti:2019,iess:2019}.

The decrease of flow velocity with depth will modify these calculations.  Suppose that at the critical depth, the characteristic velocity and turbulent length scale have decreased by a factor of 10 relative to the observed surface values.  The negative eddy viscosity would be smaller by a factor of 100, using the scaling of Sec.~\ref{sec:negative_eddy_viscosity}.  Hence, the required $\Rm_J$ falls below 1, making it marginal, while the required $\Rm_S$ is about 7, still within the regime of magnetic eddy viscosity.  The required $\eta_S = VL/\Rm_S$ is unchanged from the previous estimate.

This discussion is not intended to assert that magnetic eddy viscosity is the mechanism responsible for terminating the zonal flows in Jupiter and Saturn, but rather serves as a consideration of whether it is a plausible mechanism for gas giants at all.  There are substantial complexities not included in this discussion, such as the variation of conductivity and density with depth, meridional flow, thermal convection, and realistic physics of planetary dynamos.  Still, these estimates suggest that magnetic eddy viscosity is a potential candidate for inhibiting differential rotation in the interior of some gas giants.

\section{Conclusion}
\label{sec:conclusion}
In 2D incompressible MHD, under the assumptions of $\Rm  \gg 1$ and a turbulent flow regime where the magnetic energy is smaller than the kinetic energy, we have shown that a mean shear flow induces correlations in the magnetic field that result in a magnetic eddy viscosity.  Numerical simulations of a 2D magnetized $\beta$ plane, in which zonal flows arise naturally, confirm this result.

Two key ingredients lead directly to the prediction of magnetic eddy viscosity, given in Eqs.~\eqref{magstress_shear_relation} and \eqref{magneticdynamicviscosity}.  First is the large-$\Rm$ limit of the MHD induction equation.  Second is the assumption of turbulent flow, which acts to scramble any coherent effect built up in the magnetic field, by enforcing a short correlation time.

The expressions for the 3D generalization of the components of the Maxwell stress tensor are given in Sec.~\ref{sec:3d_generalization}.  They show directly how differential rotation leads to zonally averaged Maxwell stress that transports angular momentum.  Additionally, these expressions function as a dynamical generalization of Ferraro's law of isorotation, because they (i) allow for turbulent dynamics, (ii) do not assume axisymmetry of the magnetic field, and (iii) provide an estimate for the magnitude of the resulting force that counteracts differential rotation.

In 2D simulations with a mean poloidal magnetic field, we observe the Lorentz force transition from magnetic drag to magnetic eddy viscosity as $\Rm$ increases beyond about 10.  Magnetic eddy viscosity also occurs in simulations where the mean magnetic field is toroidal, in which case the $\omega$ effect is absent and there is no associated magnetic drag; the magnetic eddy viscosity is due solely to magnetic fluctuations.

Rotating, magnetized fluids are commonly found in astrophysical domains.  The magnetic eddy viscosity may, in certain parameter regimes, provide a simple characterization of the transport of angular momentum and the magnetic braking force due to differential rotation.  Our main numerical example here is a case where the mean shear flow consists of a turbulence-driven zonal flow, with a simple idealized model that could provide insight into the interiors of gas giants.  Mean shear flow also occurs in astrophysical disks, and magnetic eddy viscosity could be explored in that context as well.


\begin{acknowledgments}
We gratefully acknowledge useful discussions with Bill Dorland, Eli Galanti, Tristan Guillot, Yohai Kaspi, Brad Marston, Felix Parra, David Stevenson, and Steve Tobias.  The work of J.B.P. was performed under the auspices of the U.S.\ Department of Energy by Lawrence Livermore National Laboratory under Contract No.\ DE-AC52-07NA27344.  N.C.C. thanks Sean Haney for his hospitality in Barber Tract in December 2018, where part of this paper was written.  Numerical simulations were conducted on the Australian National Computational Infrastructure at ANU, which is supported by the Commonwealth of Australia. 
\end{acknowledgments}

\appendix

\section{Calculation of current and force directly from \texorpdfstring{$\v{J} \times \v{B}$}{J cross B}}	
\label{appendix:force_from_current}
Although the Maxwell stress tensor is the most direct route to the force and momentum flux derived in Sec.~\ref{sec:magneticviscosity}, we can also calculate the force by finding the current and using $\v{J} \times \v{B}$.  As in Sec.~\ref{sec:magneticviscosity}, we stick to two dimensions for simplicity.  The current is given by $\m_0 \v{J} = \v{\nabla} \times \v{B}$.  For convenience, we use somewhat different notation than before; here we refer to quantities at $t=0$ as $\v{J}_0$ and $\v{B}_0$ and refer to their first-order perturbations at $t = \D t$ as $\v{J}_1$ and $\v{B}_1$.  From the magnetic field in \eqnref{Bperturbation_shorttime}, we can compute the current,
	\begin{align}
		\m_0 \v{J}_0 &= \unit{z} \left( \pd{B_{0y}}{x} - \pd{B_{0x}}{y} \right), \\
		\m_0 \v{J}_1 &= -\unit{z} \D t \biggl[\pd{}{x} \biggl(U \pd{B_{0y}}{x} \biggr) \notag \\
			&\qquad  + \pd{}{y} \biggl( B_{0y} \pd{U}{y} - U \pd{B_{0x}}{x} \biggr) \biggr].
	\end{align}
The Lorentz force per unit volume $\v{F} = \v{J} \times \v{B}$ has an $x$ component
	\begin{equation}
		F_x = -J_z B_y.
	\end{equation}
Thus, we have
	\begin{align}
		F_{0x} &= - J_{0z} B_{0y}, \\
		F_{1x} &= -(J_{0z} B_{1y} + J_{1z} B_{0y}).
	\end{align}
We assume that $\ol{F}_{0x} = 0$, which is satisfied if $\ol{B_{0x} B_{0y}} = 0$.

Let us now consider $F_{1x}$, and substitute the previous expressions to write everything in terms of $U$ and $\v{B}_0$.  From now on, we drop the 0 subscript for simplicity.  After some algebra, we obtain for the zonally averaged force,
	\begin{equation}
		\ol{F}_{1x} = \D t \pd{}{y} \left( \frac{\ol{B_y^2}}{\m_0} \pd{U}{y} \right) + O(\D t^2).
	\end{equation}
This expression is equivalent to \eqnref{magstress_IVP}.

To summarize, we have recovered the same net force using $\v{J} \times \v{B}$ directly as we did earlier with the Maxwell stress tensor.  In the 2D MHD system where $\v{B}$ is in the $xy$ plane, both $\v{J}_0$ and $\v{J}_1$ are along $\unit{z}$, implying currents must be able to flow in the $z$ direction.

\section{Derivation of negative eddy viscosity}
\label{app:negative_eddy_viscosity}
In this appendix we derive for 2D hydrodynamic flow the negative eddy viscosity discussed in Sec.~\ref{sec:negative_eddy_viscosity} using the same technique as we used to derive the magnetic eddy viscosity.  

The zonally averaged force per unit volume in the $x$ direction is given by the Reynolds stress
	\begin{equation}
		\ol{F}_x = \pd{}{y} \ol{\r v_x v_y }.
	\end{equation}

We use the vorticity formulation of the equation of motion.  We suppose at $t=0$ there is a blob of vorticity $\z_0(x, y) = z_0 \cos(\v{k} \cdot \v{x})$.  One could multiply the spatial dependence by an envelope function to localize the spatial extent, but that is unnecessary for the effect of interest.  Consider the evolution of this vorticity blob due to advection by an imposed shear $\v{v} = U(y) \unit{x} = U_0 \sin(qy)\unit{x}$.  We neglect both the Coriolis and the Lorentz force in this derivation.   We have
	\begin{equation}
		\pd{\z}{t} = -\v{v} \cdot \v{\nabla} \z = - U_0 \sin(qy) \pd{\z}{x}.
	\end{equation}
To include the high-wave-number cutoff discussed in Sec.~\ref{sec:negative_eddy_viscosity}, one should retain the $\v{v}' \cdot \v{\nabla} \ol{\z}$ term, but we neglect it here for simplicity.

We evolve $\z$ for a short time increment $\D t$.  The vorticity becomes
	\begin{align}
		\z(\D t) &= z_0 \biggl\{ \cos(\v{k} \cdot \v{x}) \notag \\
			& \quad + \frac{\D t k_x U_0}{2} \biggl[ \cos(\v{k}_- \cdot \v{x}) - \cos(\v{k}_+ \cdot \v{x}) \biggr] \Biggr\},
	\end{align}
where $\v{k}_\pm = \v{k} \pm \v{q}$ and $\v{q} = q \unit{y}$.  From $\z = \nabla^2 \psi$, we obtain
	\begin{align}
		\psi(\D t) &= -z_0 \Biggl\{ \frac{1}{k^2} \cos(\v{k} \cdot \v{x}) \notag \\
			& + \frac{\D t k_x U_0}{2} \biggl[ \frac{1}{k_-^2} \cos(\v{k}_- \cdot \v{x}) - \frac{1}{k_+^2} \cos(\v{k}_+ \cdot \v{x}) \biggr] \Biggr\}.
	\end{align}
Calculating $\v{v}(\D t) = \unit{z} \times \v{\nabla} \psi(\D t)$ and then constructing $\ol{v_x(\D t) v_y(\D t)}$, we obtain
	\begin{equation}
		\D \ol{v_x v_y} = \frac{z_0^2 k_x^2 \D t}{2 k^4} \left( 1 - \frac{4 k_y^2}{k^2} \right) q U_0 \cos (qy) + O(\D t^2, q^3).
		\label{appendix:reynoldsstress1}
	\end{equation}
We identify $q U_0 \cos (qy) = \partial_y U$, as well as $k_x^2 z_0^2 / k^4 = k_x^2 \psi_0^2 \approx v_y^2$ and $\ol{v_y^2} = v_y^2 / 2$.  Furthermore, we observe that the factor $1 - 4k_y^2 / k^2$ is only positive for $k_x^2 > 3 k_y^2$.  This $\v{k}$-dependent prefactor does not arise in the magnetic calculation.  We collapse this factor into a $\v{k}$-independent positive constant $\g$, recognizing that $\g$ may be small in magnitude because $1 - 4k_y^2 / k^2$ is only positive in a small fraction of $\v{k}$ space.  Finally,  as in Sec.~\ref{sec:magneticviscosity}, we replace $\D t$ with $\tc$, the maximum time the shear can act coherently on a structure before being decorrelated.  Again, we interpret our results in a statistically averaged sense.  We obtain a Reynolds stress
	\begin{equation}
		\D \ol{\r v_x v_y} = \g \r \ol{v_y^2} \tc \pd{U}{y}.
	\end{equation}
This expression corresponds to a negative dynamic eddy viscosity
	\begin{equation}
		\m_e = -\g \r \ol{v_y^2}.
	\end{equation}

This calculation does not directly generalize to 3D.  Negative eddy viscosities have in the literature been limited to 2D or quasi-2D flow, and the quasi-2D nature of flow is a consequence of rotation.  In contrast, the positive magnetic eddy viscosity is derived in 3D in Section~\ref{sec:3d_generalization} without such complications.

We remark on the relation between the wave-number dependence here and that found in a related Kelvin--Orr calculation.  In that calculation with an initial wave shearing in a velocity profile $U = Sy \unit{x}$, the energy transfer to the mean flow is found to have a factor $1 - 4k_y^2/k^2$, as in Eq.~\eqref{appendix:reynoldsstress1} \cite{bakas:2013-jas,bakas:2015,constantinou:2018}.  This wave-number dependence arose when a $\v{k}$-independent friction was used as the decorrelation mechanism.  When a viscosity instead acts as the decorrelation mechanism, the length-scale dependence of the viscosity modifies the factor in the energy transfer from $1 - 4k_y^2/k^2$ to $1 - 6k_y^2/k^2$.  Returning to the calculation here, we observe that we did not include any explicit friction or viscosity.  Instead, we posited that decorrelation occurred due to turbulent inertial motion, and we implicitly assumed it was independent of wave number.  Hence, we recovered the same factor $1 - 4k_y^2/k^2$ as found using a $\v{k}$-independent friction.  

%

\end{document}